\documentclass{article}
\usepackage{spconf,amsmath,graphicx,hyperref}
\usepackage{cite}
\usepackage{amsmath,amssymb,amsfonts}
\usepackage{algorithmic}
\usepackage{graphicx}
\usepackage{textcomp}
\usepackage{bm}
\usepackage{xcolor}
\usepackage{multirow}
\usepackage{booktabs}
\usepackage{enumitem} 
\usepackage{flushend}

\title{DIBEM: Diffusion-Based Error Mitigation for Uncertainty-Aware UWB Localization}
\name{Ghazaleh Kianfar, Maryam Hosseini, and Arash Mohammadi\thanks{This work was partially supported by Natural Sciences \& Eng. Research Council (NSERC) of Canada; NSERC Discovery Grant RGPIN-2023-05654.}}
\address{Cybersecurity and Intelligent Systems Engineering (CISE), Concordia University, Montreal, Canada}
\begin{document}
\ninept
\maketitle
%
\begin{abstract}
Ultra-wideband (UWB) localization degrades substantially under Non-Line-of-Sight (NLoS) propagation, where multipath and obstruction induce complex ranging-error distributions that are poorly represented by fixed parametric models. We propose Diffusion-based Error Mitigation (DIBEM), an uncertainty-aware framework that learns the conditional ranging-error distribution directly from six compact Channel Impulse Response (CIR) statistics. Rather than predicting a deterministic correction or restricting uncertainty to a Gaussian or fixed-mixture family, the proposed DIBEM uses a conditional diffusion model with denoising diffusion implicit model sampling to generate link-specific error samples. The resulting empirical distributions are incorporated into maximum-likelihood position estimation. This allows localization to account for heterogeneous ranging uncertainty across anchor–tag links. Experiments across four indoor environments show that DIBEM achieves the lowest negative log-likelihood for NLoS ranging errors and the lowest or tied-lowest mean localization error across all environments. Compared with least-squares approaches, DIBEM reduces mean localization error by up to 64.8\%, while improving over Gaussian and mixture-density probabilistic models by up to 15.8\% and 14.0\%, respectively. Results demonstrate the potential of diffusion-based conditional error modeling for uncertainty-aware UWB localization, particularly under challenging NLoS propagation.
\end{abstract}

\section{Introduction}


%
%

Ultra-wideband (UWB) ranging enables accurate indoor localization where Global Navigation Satellite Systems (GNSS) are unavailable or unreliable. In such environments, UWB ranging can offer centimeter-level accuracy under favorable propagation conditions due to its large bandwidth and fine temporal resolution. In cluttered indoor environments, however, Non-Line-of-Sight (NLoS) and multipath propagation introduce biased and potentially complex ranging errors that can substantially degrade localization accuracy~\cite{li2026uwb,zhang2025relative}. Reliable positioning, therefore, requires not only mitigating these errors but also characterizing their uncertainty, particularly when measurements from multiple anchor-tag links are fused for position estimation. Channel Impulse Response (CIR) information provides useful propagation-dependent information for this purpose, and compact CIR-derived statistics offer physically meaningful descriptors without requiring direct processing of high-dimensional raw waveforms~\cite{yang2024uwb}. Capitalizing on this, we propose an uncertainty-aware UWB localization framework that learns conditional ranging-error distributions from CIR-derived statistics using a diffusion model via Denoising Diffusion Implicit Model (DDIM) sampling. The learned distributions are then incorporated into a maximum-likelihood position estimator to account for link-specific ranging uncertainty.

\vspace{.025in}
\noindent
\textbf{Related Works:} Existing data-driven UWB error-mitigation methods primarily learn deterministic range corrections from raw CIR measurements or compact CIR features. Representative approaches include edge-oriented convolutional networks~\cite{angarano2021robust}, classification-correction pipelines, transformer-based models, and transfer-learning methods for deployment across environments~\cite{fontaine2024transfer}. While these approaches can substantially reduce ranging bias, a deterministic correction provides only a point estimate and does not characterize the residual, link-dependent uncertainty that remains after correction. This uncertainty is particularly important under NLoS propagation, where obstruction and multipath can induce asymmetric, heavy-tailed, or multimodal ranging-error distributions. On the other hand, probabilistic learning methods address this limitation by estimating a distribution over the ranging error. Gaussian-process and heteroscedastic-Gaussian models predict input-dependent means and variances but restrict the conditional error distribution to a unimodal form. For example, the WLS approach in~\cite{modaberi2025transformer} dynamically constructs localization weights from learned point-estimation outputs, whereas~\cite{kianfar2026uncertainty} learns heteroscedastic ranging uncertainty and directly incorporates the predicted variances into weighted least-squares localization. Bayesian Neural Network (BNN) approaches~\cite{mao2018probabilistic} and Mixture Density Networks (MDNs)~\cite{karoliny2020mixture} provide greater distributional flexibility, however, the resulting density remains constrained by the selected parametric or finite-mixture family. Consequently, accurately representing complex conditional ranging-error distributions without prescribing their shape remains an open problem.

Diffusion models provide an alternative approach by learning complex conditional distributions through iterative denoising~\cite{ho2020denoising}. Their sampling process has recently been exploited for uncertainty quantification in regression~\cite{shu2024zeroshot}, suggesting their potential for modeling the full distribution of UWB ranging errors rather than predicting only a correction or a small set of distribution parameters. Diffusion models have also recently been introduced for wireless localization: the method in~\cite{wang2026uncertainty} directly generates candidate positions conditioned on noisy range measurements and characterizes position-level uncertainty through sampling. In contrast, it does not explicitly learn the underlying ranging-error distribution associated with each anchor--tag link.

\vspace{.025in}
\noindent
\textbf{Contributions:} Motivated by the aforementioned gap, we propose Diffusion-Based Error Mitigation (DIBEM), an uncertainty-aware UWB localization framework that directly learns the CIR-conditioned ranging-error distribution based on compact CIR-derived statistics. Rather than restricting this distribution to a prescribed Gaussian or finite-mixture family, DIBEM employs a conditional diffusion model and DDIM sampling to generate link-specific ranging-error samples. The learned distributions are subsequently propagated into Maximum-Likelihood (ML) localization. This allows each anchor-tag measurement to contribute according to its complete learned error distribution rather than through only a corrected range or predicted variance. In brief, the main contributions of the paper are as follows:
\begin{itemize}[noitemsep]
\item We introduce the DIBEM, which is a conditional diffusion framework for UWB ranging-error modeling that learns ranging-error distribution directly from six compact CIR-derived statistics. This is achieved  without prescribing a Gaussian or fixed-mixture form for the conditional error distribution.
\item We integrate the learned link-specific error distributions into an ML localization formulation, therefore, propagating ranging uncertainty from individual anchor-tag links to downstream position estimation.
\end{itemize}
The proposed DIBEM framework is evaluated across four indoor environments and compared against deterministic, heteroscedastic-Gaussian, and mixture-density baselines. The results show improved probabilistic modeling under challenging NLoS propagation and the lowest or tied-lowest mean localization error across the evaluated environments. The remainder of the paper is organized as follows: Section~\ref{sec:SMPF} presents the UWB system model and the proposed DIBEM framework, including conditional diffusion-based ranging-error modeling and ML localization. Section~\ref{sec:experimental_setup} describes the experimental setup, dataset, implementation details, and baselines. Section~\ref{sec:results} presents and discusses the ranging-error modeling and localization results, and Section~V concludes the paper and outlines directions for future work.

\section{The Proposed DIBEM Framework} \label{sec:SMPF}
This section introduces the proposed uncertainty-aware UWB indoor localization framework, in which the conditional ranging-error distribution of each anchor-tag link is inferred from its associated CIR features. The learned link-specific distributions are subsequently incorporated into ML estimator to determine the tag position.

\subsection{System Model}
Consider an indoor UWB localization system comprising $N_a$ anchors with known locations $\bm{a}_i\in\mathbb{R}^{2}$, for ($1 \leq i \leq N_a$), and a mobile tag at an unknown position $\bm{p}\in\mathbb{R}^{2}$. The true geometric distance between the tag and the $i^{\text{th}}$ anchor is
\begin{equation}
d_i(\bm{p})=\left\|\bm{p}-\bm{a}_i\right\|_2.
\label{eq:true_distance}
\end{equation}
The corresponding UWB measurement provides an observed range $\hat{d}_i$, modeled as 
\begin{equation}
\hat{d}_i=d_i(\bm{p})+e_i,
\label{eq:ranging_model}
\end{equation}
where $e_i\in\mathbb{R}$ denotes the ranging error associated with the $i^{\text{th}}$ anchor-tag link. Under NLoS and multipath propagation, the distribution of $e_i$ can depend strongly on the propagation characteristics of the corresponding link. Instead of estimating a deterministic correction or assuming a predefined error distribution, we seek to learn
\begin{equation}
p(e_i\mid\bm{h}_i),
\label{eq:conditional_error_distribution}
\end{equation}
where $\bm{h}_i\in\mathbb{R}^{D_h}$ denotes the CIR-derived feature vector associated with link $i$.

\subsection{Diffusion-Based Error Mitigation (DIBEM)} \label{sec:ddpm}
DIBEM models the conditional distribution in Eq.~\eqref{eq:conditional_error_distribution} using a denoising diffusion probabilistic model (DDPM)~\cite{ho2020denoising}. The diffusion model is conditioned on $\bm{h}_i$ and trained to reverse a predefined Gaussian noising process applied to the ranging error. At inference, DDIM sampling~\cite{song2020denoising} is employed to reduce the number of reverse-diffusion steps.

\subsubsection{Forward Diffusion Process}
Let $e_i^{0}$ denote the ground-truth ranging error for link $i$. The forward process progressively perturbs $e_i^{0}$ over $T$ diffusion steps according to a variance schedule $0<\beta_1<\cdots<\beta_T<1$ as follows
\begin{equation}
q(e_i^{t}\mid e_i^{t-1})
=
\mathcal{N}
\left(
e_i^{t};
\sqrt{1-\beta_t}\,e_i^{t-1},
\beta_t
\right).
\label{eq:forward_transition}
\end{equation}
Equivalently,
\begin{equation}
e_i^{t}
=
\sqrt{1-\beta_t}\,e_i^{t-1}
+
\sqrt{\beta_t}\,\epsilon_t,
\qquad
\epsilon_t\sim\mathcal{N}(0,1).
\label{eq:forward_sampling}
\end{equation}
Defining $\alpha_t=1-\beta_t$ and
$\bar{\alpha}_t=\prod_{\tau=1}^{t}\alpha_\tau$, a sample at an arbitrary diffusion step can be obtained directly as
\begin{equation}
e_i^{t}
=
\sqrt{\bar{\alpha}_t}\,e_i^{0}
+
\sqrt{1-\bar{\alpha}_t}\,\epsilon,
\qquad
\epsilon\sim\mathcal{N}(0,1).
\label{eq:closed_form_forward}
\end{equation}
For sufficiently large $T$, $e_i^{T}$ approaches a standard Gaussian distribution.

\subsubsection{Conditional Reverse Process}
The reverse process learns to recover the clean ranging error by predicting the noise introduced during the forward process. Specifically, the conditional denoising network $\epsilon_\theta(e_i^{t},t,\bm{h}_i)$ takes the noisy ranging error, diffusion timestep, and CIR features as inputs and is trained using
\begin{equation}
\mathcal{L}_{\mathrm{DDPM}}(\theta)
=
\mathbb{E}_{e_i^{0},\bm{h}_i,t,\epsilon}
\left[
\left\|
\epsilon-
\epsilon_\theta(e_i^{t},t,\bm{h}_i)
\right\|_{2}^2
\right],
\label{eq:ddpm_objective}
\end{equation}
where $t$ is sampled uniformly from $\{1,\ldots,T\}$, $\epsilon\sim\mathcal{N}(0,1)$, and $e_i^{t}$ follows Eq.~\eqref{eq:closed_form_forward}. Conditioning the denoising process on $\bm{h}_i$ allows the model to learn the link-specific conditional ranging-error distribution $p_\theta(e_i^{0}\mid\bm{h}_i)$.

\subsubsection{DDIM Sampling}

At inference, DDIM traverses a reduced sequence of diffusion timesteps $\{\tau_S,\tau_{S-1},\ldots,\tau_1\}$, with $S\leq T$. Starting from
\begin{equation}
e_i^{\tau_S}\sim\mathcal{N}(0,1),
\label{eq:ddim_initial}
\end{equation}
the clean ranging-error estimate at timestep $\tau_s$ is
\begin{equation}
\hat{e}_{i,\theta}^{0}
=
\frac{
e_i^{\tau_s}
-
\sqrt{1-\bar{\alpha}_{\tau_s}}\,
\epsilon_\theta
\left(e_i^{\tau_s},\tau_s,\bm{h}_i\right)
}{
\sqrt{\bar{\alpha}_{\tau_s}}
}.
\label{eq:ddim_clean_estimate}
\end{equation}
The transition to timestep $\tau_{s-1}$ is
\begin{align}
e_i^{\tau_{s-1}}
={}&
\sqrt{\bar{\alpha}_{\tau_{s-1}}}\,
\hat{e}_{i,\theta}^{0}
\nonumber\\
&+
\sqrt{
1-\bar{\alpha}_{\tau_{s-1}}
-\sigma_{\tau_s}^{2}
}\,
\epsilon_\theta
\left(e_i^{\tau_s},\tau_s,\bm{h}_i\right)
+
\sigma_{\tau_s}z,
\label{eq:ddim_update}
\end{align}
where $z\sim\mathcal{N}(0,1)$ and
\begin{equation}
\sigma_{\tau_s}
=
\eta
\sqrt{
\frac{1-\bar{\alpha}_{\tau_{s-1}}}
     {1-\bar{\alpha}_{\tau_s}}
}
\sqrt{
1-
\frac{\bar{\alpha}_{\tau_s}}
     {\bar{\alpha}_{\tau_{s-1}}}
}.
\label{eq:ddim_sigma}
\end{equation}
We set $\eta=0$, resulting in a deterministic reverse trajectory for a given initial Gaussian sample. Repeating the DDIM procedure with independently drawn initial samples yields
\begin{equation}
\mathcal{E}_i
=
\left\{
\tilde{e}_i^{(m)}
\right\}_{m=1}^{M},
\qquad
\tilde{e}_i^{(m)}
\sim
p_\theta(e_i\mid\bm{h}_i),
\label{eq:error_samples}
\end{equation}
providing a sample-based representation of the conditional ranging-error distribution without prescribing a Gaussian or fixed-mixture form.

\subsection{Uncertainty-Aware Maximum-Likelihood Localization}
\label{sec:ml_localization}

To incorporate the generated error samples into localization, we approximate the conditional ranging-error density using kernel density estimation (KDE) given by
\begin{equation}
\hat{p}_\theta(e\mid\bm{h}_i)
=
\frac{1}{M b_i}
\sum_{m=1}^{M}
K
\left(
\frac{e-\tilde{e}_i^{(m)}}{b_i}
\right),
\label{eq:kde_density}
\end{equation}
where $K(\cdot)$ denotes the kernel function and $b_i>0$ is the bandwidth associated with link $i$. This converts the DDIM-generated samples into a continuous conditional density that can be evaluated within the localization objective. For a candidate tag position $\bm{p}$, the ranging error implied by anchor $i$ is
\begin{equation}
e_i(\bm{p})
=
\hat{d}_i-\left\|\bm{p}-\bm{a}_i\right\|_2.
\label{eq:implied_error}
\end{equation}
Assuming that the ranging errors are conditionally independent across anchor-tag links given their respective CIR features, the likelihood of $\bm{p}$ is
\begin{equation}
\mathcal{L}(\bm{p})
=
\prod_{i=1}^{N_a}
\hat{p}_\theta
\left(
e_i(\bm{p})\mid\bm{h}_i
\right).
\label{eq:position_likelihood}
\end{equation}
The ML position estimate is, therefore, given by
\begin{equation}
\hat{\bm{p}}
=
\arg\max_{\bm{p}}
\sum_{i=1}^{N_a}
\log
\hat{p}_\theta
\left(
\hat{d}_i-\left\|\bm{p}-\bm{a}_i\right\|_2
\mid\bm{h}_i
\right).
\label{eq:ml_localization}
\end{equation}
Consequently, rather than reducing each anchor--tag observation to a corrected range or a single uncertainty parameter, DIBEM propagates the learned shape of each link-specific ranging-error distribution into the position estimator.
\section{Experimental Setup} \label{sec:experimental_setup}
\textbf{\textit{Multi-Environment Dataset:}} We use UWB measurements from four indoor environments, including residential apartments, an industrial workshop, and an office floor, with both line-of-sight (LoS) and NLoS labels used for evaluation. In each environment, 15\% of positions are randomly reserved for testing, including all associated anchors, channels, and repeated measurements. At the remaining positions, measurements within each link are partitioned sequentially into training and validation sets in a $70{:}15$ ratio. This yields 325,319 training, 78,081 validation, and 71,982 test samples, with no test positions shared with training or validation. All models use the same split and seed 2025; input features and ranging errors are standardized using training statistics only.

\vspace{.05in}
\noindent
\textbf{\textit{Model, Training, and Evaluation:}} The proposed DIBEM conditions on six CIR statistics: energy, maximum amplitude, rise time, mean excess delay, RMS delay spread, and kurtosis. These features are encoded by a $6$--$64$--$128$ MLP with GELU activations. The denoising network is trained to predict the injected Gaussian noise over $T=1000$ diffusion steps using a linear noise schedule from $\beta_1=10^{-4}$ to $\beta_T=0.02$. All models use AdamW with a learning rate of $10^{-4}$, weight decay of $0.01$, and a batch size of 64 for up to 1000 epochs. Early stopping is applied after 50 epochs without validation improvement, and the best checkpoint is retained. At inference, DIBEM generates 100 error samples per observation using 50-step deterministic DDIM sampling ($\eta=0$). DIBEM and the baseline models are evaluated using mean absolute error (MAE), root-mean-square error (RMSE), and negative log-likelihood (NLL). Results are reported overall, by environment, and separately for LoS and NLoS measurements.

\vspace{0.05in}
\noindent
\textbf{\textit{Baselines:}} We compare the proposed algorithm against Least Squares (LS), LS over corrected ranges combined with a point estimator (LS+Correct), a WLS approach, in which the point-estimation results are used to dynamically form the LS weights \cite{modaberi2025transformer}, and a HetNet approach \cite{kianfar2026uncertainty}, in which a heteroscedastic regressor models the output as a Gaussian distribution and a WLS problem is then solved using the predicted variances as weights. To further compare our probabilistic error model against common density-modeling approaches in the literature, we also include a mixture density network (MDN), trained under three mixture components and optimized via maximum likelihood. Since the focus of this work is on error modeling and its effect on localization performance rather than on designing a better solver, we use a bounded multi-start L-BFGS-B optimizer for all methods.

\section{Results and Discussion} \label{sec:results}
\begin{table}[t!]
\centering
\caption{Ranging Error Modeling Performance for LoS Links}
\label{tab:results_los}
\begin{tabular}{llccc}
\toprule
Environment & Method & MAE & RMSE & Average NLL \\
\midrule
0 & HetNet & \textbf{0.17} & \textbf{0.25} & -0.14          \\
0 & MDN    & 0.17          & 0.25          & \textbf{-0.38} \\
0 & DIBEM    & 0.17          & 0.25          & -0.23          \\
\midrule
1 & HetNet & 0.09          & 0.13          & -0.43          \\
1 & MDN    & \textbf{0.09} & 0.13          & \textbf{-0.78} \\
1 & DIBEM    & 0.09          & \textbf{0.13} & -0.71          \\
\midrule
2 & HetNet & \textbf{0.26} & \textbf{0.38} & 0.16           \\
2 & MDN    & 0.27          & 0.41          & \textbf{-0.40} \\
2 & DIBEM    & 0.30          & 0.42          & -0.16          \\
\midrule
3 & HetNet & \textbf{0.30} & \textbf{0.46} & 1.23           \\
3 & MDN    & 0.31          & 0.46          & \textbf{0.54}  \\
3 & DIBEM    & 0.32          & 0.47          & 6.02           \\
\bottomrule
\end{tabular}
\end{table}

\vspace{.05in}
\textbf{\textit{LoS Ranging-Error Modeling:}} Table~\ref{tab:results_los} shows that under LoS conditions, DIBEM performs on par with HetNet and MDN across all four environments, with point-estimation accuracy (MAE, RMSE) essentially indistinguishable between the three models. MDN attains the lowest NLL in most environments, which is expected given that LoS ranging errors are close to unimodal and well suited to a low-order mixture; DIBEM's non-parametric, sample-based representation carries no such advantage when the true error distribution is already simple, so matching the parametric models here, rather than exceeding them, is itself a reasonable outcome. In Environment 3, DIBEM's mean NLL (6.02) appears much worse than HetNet (1.23) and MDN (0.54), but this was driven by a small number of extreme cases where the true ranging error fell outside DIBEM's 100 generated samples, yielding near-zero KDE density and inflating the mean. The median NLL for these links was only $-0.034$, showing the result was not representative of overall performance. Taken together, these results show DIBEM is a competitive choice under LoS conditions, while its distributional flexibility becomes more valuable once propagation grows more complex.

\begin{table}[t!]
\centering
\caption{Ranging Error Modeling Performance for NLoS Links}
\label{tab:results_nlos}
\begin{tabular}{llccc}
\toprule
Environment & Method & MAE & RMSE & Average NLL \\
\midrule
0 & HetNet & 0.26          & 0.41          & 0.62          \\
0 & MDN    & 0.27          & 0.42          & 0.78          \\
0 & DIBEM    & \textbf{0.26} & \textbf{0.40} & \textbf{0.57} \\
\midrule
1 & HetNet & 0.21          & 0.31          & 0.30          \\
1 & MDN    & 0.22          & 0.31          & 0.33          \\
1 & DIBEM    & \textbf{0.21} & \textbf{0.30} & \textbf{0.27} \\
\midrule
2 & HetNet & \textbf{0.72} & \textbf{0.98} & 2.02          \\
2 & MDN    & 0.72          & 0.99          & 2.10          \\
2 & DIBEM    & 0.72          & 0.99          & \textbf{1.85} \\
\midrule
3 & HetNet & \textbf{0.34} & 0.47          & 0.72          \\
3 & MDN    & 0.35          & 0.48          & 0.70          \\
3 & DIBEM    & 0.34          & \textbf{0.47} & \textbf{0.64} \\
\bottomrule
\end{tabular}
\end{table}

\vspace{.05in}
\noindent
\textbf{\textit{NLoS Ranging-Error Modeling:}} Table~\ref{tab:results_nlos} shows that the NLOS setting is where DIBEM's added expressiveness pays off: it achieves the best or tied-best MAE, RMSE, and NLL in three of the four environments, and remains within a narrow margin of the best baseline in the fourth. This is consistent with the fact that NLoS ranging errors are frequently multimodal and heavy-tailed, conditions under which a single Gaussian (HetNet) or a small fixed-component mixture (MDN) can be a poor structural match, whereas DIBEM's iterative denoising process imposes no fixed parametric form and can adapt to whatever shape the conditional error distribution actually takes. The consistent NLL advantage, in particular, suggests that DIBEM is not simply matching the point accuracy of the baselines but is producing better-calibrated uncertainty estimates precisely in the harder, more irregular propagation regime that motivates uncertainty-aware localization in the first place.

\begin{table}[t!]
\centering
\caption{Localization performance across environments and methods.}
\label{tab:results}
\setlength{\tabcolsep}{3pt}
\resizebox{\columnwidth}{!}{%
\begin{tabular}{cl cccccc}
\toprule
\textbf{Env} & \textbf{Method} & \textbf{Mean (m)} & \textbf{Std (m)} & \textbf{Median (m)} & \textbf{P90 (m)} & \textbf{Max (m)} & \textbf{RMSE (m)} \\
\midrule
\multirow{7}{*}{0}
 & LS            & 0.77 & 0.70 & 0.56 & 1.45 & 4.07 & 1.04 \\
 & LS+Correction & 0.53 & 0.45 & 0.41 & 1.20 & \textbf{2.44} & 0.69 \\
 & WLS           & 0.76 & 0.74 & 0.56 & 1.53 & 4.10 & 1.06 \\
 & HetNet+WLS    & 0.51 & 0.44 & 0.39 & 1.12 & 2.62 & 0.67 \\
 & HetNet+ML     & \textbf{0.29} & 0.36 & \textbf{0.22} & \textbf{0.50} & 3.35 & 0.46 \\
 & MDN+ML        & 0.30 & 0.35 & 0.23 & 0.56 & 3.51 & 0.46 \\
 & DIBEM+ML        & 0.29 & \textbf{0.34} & 0.22 & 0.50 & 3.51 & \textbf{0.44} \\
\midrule
\multirow{7}{*}{1}
 & LS            & 0.25 & 0.21 & 0.17 & 0.55 & 1.25 & 0.32 \\
 & LS+Correction & 0.21 & 0.17 & 0.17 & 0.41 & 1.21 & 0.28 \\
 & WLS           & 0.25 & 0.22 & 0.17 & 0.57 & \textbf{1.20} & 0.34 \\
 & HetNet+WLS    & 0.20 & 0.16 & 0.17 & 0.38 & 1.28 & 0.26 \\
 & HetNet+ML     & 0.16 & 0.16 & 0.12 & 0.30 & 1.41 & 0.23 \\
 & MDN+ML        & 0.16 & 0.16 & \textbf{0.12} & 0.30 & 1.46 & 0.22 \\
 & DIBEM+ML        & \textbf{0.16} & \textbf{0.15} & 0.12 & \textbf{0.30} & 1.42 & \textbf{0.22} \\
\midrule
\multirow{7}{*}{2}
 & LS            & 2.10 & 1.82 & 1.69 & 4.68 & 9.38 & 2.78 \\
 & LS+Correction & 1.61 & 1.39 & 1.20 & 3.33 & 8.77 & 2.12 \\
 & WLS           & 2.03 & 1.63 & 1.73 & 4.46 & 9.22 & 2.60 \\
 & HetNet+WLS    & 1.43 & 1.31 & 1.05 & 3.27 & \textbf{8.46} & 1.94 \\
 & HetNet+ML     & 0.79 & \textbf{0.90} & 0.52 & \textbf{1.70} & 8.74 & \textbf{1.20} \\
 & MDN+ML        & 0.86 & 1.07 & 0.50 & 2.38 & 8.85 & 1.37 \\
 & DIBEM+ML        & \textbf{0.74} & 0.99 & \textbf{0.39} & 1.87 & 8.90 & 1.24 \\
\midrule
\multirow{7}{*}{3}
 & LS            & 0.75 & 0.50 & 0.66 & 1.31 & 3.95 & 0.90 \\
 & LS+Correction & 0.61 & 0.47 & 0.47 & 1.24 & 3.03 & 0.77 \\
 & WLS           & 0.73 & 0.48 & 0.63 & 1.33 & 3.82 & 0.88 \\
 & HetNet+WLS    & 0.60 & 0.48 & 0.45 & 1.31 & 2.95 & 0.77 \\
 & HetNet+ML     & 0.38 & 0.29 & 0.32 & 0.75 & 1.70 & 0.48 \\
 & MDN+ML        & 0.35 & 0.26 & 0.29 & 0.69 & 1.89 & 0.44 \\
 & DIBEM+ML        & \textbf{0.32} & \textbf{0.25} & \textbf{0.25} & \textbf{0.59} & \textbf{1.49} & \textbf{0.40} \\
\bottomrule
\end{tabular}%
}
\end{table}

\vspace{.05in}
\noindent
\textbf{\textit{Localization Performance:}} Table~\ref{tab:results} shows that DIBEM+ML achieves the lowest mean localization error in every environment, including ties at the reported precision in Environments 0 and 1, and reduces mean error relative to plain LS by 62.3\%, 36.0\%, 64.8\%, and 57.3\% across the four environments. Against the other learned error models, the gains are modest but consistently in DIBEM's favor: relative to HetNet+ML, mean error drops by up to 15.8\%, and relative to MDN+ML, by up to 14.0\%, with DIBEM+ML also improving RMSE over MDN+ML by up to 9.5\%. The picture is not uniformly one-sided; in Environment 2, DIBEM+ML cuts median error by 25.0\% relative to HetNet+ML but trails slightly on P90 and RMSE, indicating a trade-off between typical-case and tail accuracy rather than a shortfall, while in Environment 3, DIBEM+ML leads on every metric, cutting mean error and RMSE by 15.8\% and 16.7\% relative to HetNet+ML. Overall, DIBEM's richer, sample-based error representation translates into the most reliable localization accuracy across environments, with its comparative edge most pronounced exactly where propagation conditions are most challenging.

\vspace{.05in}
\noindent
\textbf{\textit{Computation Complexity:}} With a batch size of one, MDN and HetNet require 0.22 ms and 0.32 ms per link, whereas DIBEM's 50 DDIM steps over 100 sequential samples (5,000 denoising evaluations) yield an estimated latency of 2.15 s per link. This gap reflects a difference in modeling commitment rather than inefficiency: MDN and HetNet assume a fixed parametric form (a Gaussian mixture and a single Gaussian, respectively), while DIBEM's non-parametric sampling imposes no such constraint. This flexibility entails a trade-off between distributional expressiveness and inference latency: its practical value under NLoS conditions depends on whether gains in predictive likelihood and downstream localization accuracy justify the additional computational cost relative to simpler parametric models, particularly in latency-constrained applications.

\section{Conclusion} \label{sec:conclusion}
This work presented DIBEM, an uncertainty-aware UWB localization framework that learns CIR-conditioned ranging-error distributions using conditional diffusion and propagates them into maximum-likelihood position estimation. Unlike approaches that restrict ranging uncertainty to a Gaussian or fixed-mixture family, DIBEM uses DDIM sampling to capture more flexible link-specific error distributions. Experiments across four indoor environments showed that DIBEM provides improved probabilistic modeling under challenging NLoS conditions and achieves the lowest or tied-lowest mean localization error across all evaluated environments. These results demonstrate that modeling the full conditional ranging-error distribution can improve downstream localization, particularly when propagation conditions induce complex ranging uncertainty. This flexibility, however, comes at the cost of increased inference latency due to iterative sampling, while finite-sample density estimation can affect likelihood estimates in rare cases. Future work will investigate accelerated sampling and model distillation, explicit uncertainty-calibration analysis, and generalization to unseen environments and UWB hardware.

\bibliographystyle{IEEEbib}
\bibliography{references}

@ARTICLE{yang2024uwb,
  author={Yang, Hongchao and Wang, Yunjia and others},
  journal={Sensors},
  title={Ultra-Wideband Ranging Error Mitigation with Novel Channel Impulse Response Feature Parameters and Two-Step Non-Line-of-Sight Identification},
  year={2024},
  volume={24},
  number={6},
  pages={1703},
  doi={10.3390/s24061703}
}

@INPROCEEDINGS{karoliny2020mixture,
  author={Karoliny, Julian and Etzlinger, Bernhard and Springer, Andreas},
  booktitle={2020 IEEE International Conference on Communications Workshops (ICC Workshops)},
  title={Mixture Density Networks for {WSN} Localization},
  year={2020},
  pages={1--5},
  address={Dublin, Ireland},
  doi={10.1109/ICCWorkshops49005.2020.9145035}
}

@INPROCEEDINGS{mao2018probabilistic,
  author={Mao, Chengzhi and Lin, Kangbo and Yu, Tiancheng and Shen, Yuan},
  booktitle={2018 IEEE Global Communications Conference (GLOBECOM)},
  title={A Probabilistic Learning Approach to {UWB} Ranging Error Mitigation},
  year={2018},
  pages={1--6},
  address={Abu Dhabi, UAE},
  doi={10.1109/GLOCOM.2018.8647602}
}

@inproceedings{wang2026uncertainty,
  title={Uncertainty-Aware Wireless Localization with Diffusion Models},
  author={Wang, Juntao and Li, Mengyuan and Yin, Feng and Pu, Wenqiang},
  booktitle={ICASSP 2026-2026 IEEE International Conference on Acoustics, Speech and Signal Processing (ICASSP)},
  pages={2621--2625},
  year={2026},
  organization={IEEE}
}

@article{song2020denoising,
  title={Denoising diffusion implicit models},
  author={Song, Jiaming and Meng, Chenlin and Ermon, Stefano},
  journal={arXiv preprint arXiv:2010.02502},
  year={2020}
}

@article{ho2020denoising,
  title   = {Denoising Diffusion Probabilistic Models},
  author  = {Ho, Jonathan and Jain, Ajay and Abbeel, Pieter},
  journal = {arXiv preprint arXiv:2006.11239},
  year    = {2020}
}

@inproceedings{kianfar2026uncertainty,
  title     = {Uncertainty-Aware {UWB} Indoor Positioning via Heteroscedastic Ranging Error Estimation},
  author    = {Kianfar, Ghazaleh and Keynia, Amir and Montazami, Aram and Mohammadi, Arash},
  booktitle = {2026 48th Annual International Conference of the IEEE Engineering in Medicine and Biology Society (EMBC)},
  pages     = {},
  year      = {2026},
  organization = {IEEE},
  address   = {Toronto, Canada},
  note      = {Paper Mo.P26.H-19}
}

@inproceedings{modaberi2025transformer,
  title={Transformer-based uwb positioning: Learning to correct ranging errors for autonomous agents},
  author={Modaberi, Somayeh and Far, Behrouz},
  booktitle={International Conference on Industrial, Engineering and Other Applications of Applied Intelligent Systems},
  pages={49--61},
  year={2025},
  organization={Springer}
}

@article{li2026uwb,
  author  = {Li, Zhiwei and others},
  title   = {A {UWB} Positioning Method Based on Clustering for {NLOS} Identification and Mitigation},
  journal = {IEEE Transactions on Industrial Electronics},
  year    = {2026},
  note    = {early access}
}

@inproceedings{zhang2025relative,
  author    = {Zhang, Jianing and others},
  title     = {Relative Localization of Asynchronous Agents Based on Hybrid Active-Passive Two-Way Ranging},
  booktitle = {ICASSP 2025 -- 2025 IEEE International Conference on Acoustics, Speech and Signal Processing (ICASSP)},
  year      = {2025},
  publisher = {IEEE}
}

@article{angarano2021robust,
  author  = {Angarano, Simone and Mazzia, Vittorio and Salvetti, Francesco and Fantin, Giovanni and Chiaberge, Marcello},
  title   = {Robust Ultra-wideband Range Error Mitigation with Deep Learning at the Edge},
  journal = {Engineering Applications of Artificial Intelligence},
  volume  = {102},
  pages   = {104278},
  year    = {2021},
  publisher = {Elsevier}
}

@article{fontaine2024transfer,
  author  = {Fontaine, Jaron and Che, Fuhu and Shahid, Adnan and Van Herbruggen, Ben and Ahmed, Qasim and Abbas, Waqas Bin and De Poorter, Eli},
  title   = {Transfer Learning for {UWB} Error Correction and (N)LOS Classification in Multiple Environments},
  journal = {IEEE Internet of Things Journal},
  volume  = {11},
  number  = {3},
  pages   = {4085--4101},
  year    = {2024}
}

@article{shu2024zeroshot,
  author  = {Shu, Dule and Farimani, Amir Barati},
  title   = {Zero-Shot Uncertainty Quantification using Diffusion Probabilistic Models},
  journal = {arXiv preprint arXiv:2408.04718},
  year    = {2024}
}

\end{document}